\documentclass[9pt,twocolumn,twoside]{osajnl}

\journal{ol} 

\setboolean{shortarticle}{true}

\usepackage{xcolor}
\usepackage{siunitx}
\usepackage{graphicx}
\usepackage{amsmath}
\usepackage{textcomp}
\usepackage{listings}
\usepackage{courier}
\usepackage{booktabs}
\usepackage{nicefrac, xfrac}

\graphicspath{{img/}}

\newcommand{\pump}{\SI{1064}{nm}}
\newcommand{\convert}{\SI{2128}{nm}}

\usepackage{wrapfig}
\usepackage{graphicx}
\usepackage{lipsum}

\usepackage{todonotes}
\presetkeys{todonotes}{inline,inlinewidth=4cm}{}

\begin{document}

\title{30 W ultra-stable laser light at 2128 nm for future gravitational-wave observatories}

\author[1]{Julian Gurs}
\author[2]{Nina Bode}
\author[3]{Christian Darsow-Fromm}
\author[2]{Henning Vahlbruch}
\author[1]{Pascal Gewecke}
\author[4,5]{Sebastian Steinlechner}
\author[2]{Benno Willke}
\author[1]{Roman Schnabel}

\affil[1]{Institut für Quantenphysik, Universität Hamburg, Luruper Chaussee 149, 22761 Hamburg, Germany}
\affil[2]{Institut für Gravitationsphysik, Leibniz Universität Hannover and Max-Planck-Institut für Gravitationsphysik (Albert-Einstein-Institut), Callinstr. 38, 30167 Hannover, Germany}
\affil[3]{Institut für Experimentalphysik, Universität Hamburg , Luruper Chaussee 149, 22761 Hamburg, Germany}
\affil[4]{Faculty of Science and Engineering, Maastricht University, Duboisdomein 30, 6229 GT Maastricht, The Netherlands}
\affil[5]{Nikhef, Science Park 105, 1098 XG Amsterdam, The Netherlands}

\begin{abstract}
    Thermal noise of the dielectric mirror coatings can limit laser-optical high-precision measurements. 
Coatings made of amorphous silicon and silicon nitride could provide a remedy for both gravitational-wave (GW) detectors and optical clocks.
However, the absorption spectra of these materials require laser wavelengths around 2\,µm. For GW detectors, ultra-stable laser light of tens or hundreds of watts is needed. 
Here, we report the production of nearly 30\,W of ultra-stable laser light at 2128\,nm by frequency conversion of 1064\,nm light from a master oscillator power amplifier system.
We achieve an external conversion efficiency of (67.5\,±\,0.5)\,\% via optical parametric oscillation and a relative power noise in the range of $\boldmath\mathbf{10^{-6}/\sqrt{\textbf{Hz}}}$ at 100\,Hz, which is almost as low as that of the input light and underlines the potential of our approach.%
\end{abstract}

\maketitle

\section{Introduction}
\label{sec:introduction}
The first breakthroughs in gravitational-wave detection, the observations of merging black holes \cite{abbottObservationGravitationalWaves2016} and of binary neutron star coalescences \cite{ligoscientificcollaborationandvirgocollaborationGW170817ObservationGravitational2017a}, by the Advanced LIGO \cite{Aasi_2015} and Advanced Virgo \cite{Acernese_2015} GW detectors, gave birth to gravitational-wave astronomy \cite{theligoscientificcollaboration2023gwtc3}. 
A further sensitivity increase of the existing detectors network, which also includes GEO600 \cite{Dooley_2016} and KAGRA \cite{KAGRA}, will give access to other astrophysical and even cosmological sources.\\ 
Proposed new upgrades and observatories, such as the Einstein Telescope \cite{ETSteeringCommitteeEditorialTeam2020}, LIGO Voyager \cite{adhikariCryogenicSiliconInterferometer2020}, and Cosmic Explorer \cite{reitze2019cosmic}, demand further improvements of existing and development of new approaches to further reduce fundamental noise sources.
For example, thermally driven noise of the mirror coating material shall be reduced by lowering the temperature of the test-mass mirrors. 
However, with the current test-mass material, fused silica, a large increase in the mechanical loss of the substrates at low temperatures cancels out the advantage of cryogenic operation~\cite{Nawrodt2011}.
A promising candidate to avoid these issue are crystalline silicon test masses, potentially in conjunction with crystalline AlGaAs coatings~\cite{Cole2023,Kedar_23}. 
Because of the transparency window of silicon, the operating wavelength would have to be increased to at least \SI{1.5}{\micro\metre}. 
Suitable \SI{1.55}{\micro\metre} laser sources exceeding \SI{100}{\watt} output power and fulfilling the noise requirements of GW detectors have been demonstrated \cite{DeVarona2017}. 
Another coating option would be based on amorphous silicon and silicon nitride \cite{steinlechnerSiliconBasedOpticalMirror2018a}. 
Interest in those coatings has recently increased because excess noise was found in crystalline AlGaAs coatings \cite{Kedar_23}.  
To get the full potential of this new material combination, the operation wavelength of the interferometer needs to be changed to a wavelength around \SI{2}{\micro\metre}~\cite{adhikariCryogenicSiliconInterferometer2020}.
In this wavelength regime, laser sources based on diode lasers, Ho:YAG, Ho:YLF, Tm:YAG, Tm:YAP, Tm:LiYF$_4$, and Ho:GdVO$_4$ \cite{Kapasi:20,Tian2023,Bollig:09,soton30180,Zhou:23,Alles2023,Wu2023} have been developed, showing high continuous-wave output powers \cite{Zhang:20} and compatibility with advanced quantum-measurement techniques such as squeezed states of light \cite{mansellObservationSqueezedLight2018,yapSqueezedVacuumPhase2019}.
Another proposed wavelength is \convert{}, which can be obtained by wavelength-doubling of the existing \pump{} laser systems \cite{Darsow-Fromm2020}. 
This technique should be able to directly transfer the high stability and low-noise qualities of laser systems that have been developed for GW detectors, and is immediately extendable for the generation of squeezed light \cite{Darsow-Fromm2021}.
\begin{figure*}
	\centering
	\includegraphics[width=0.8\linewidth]{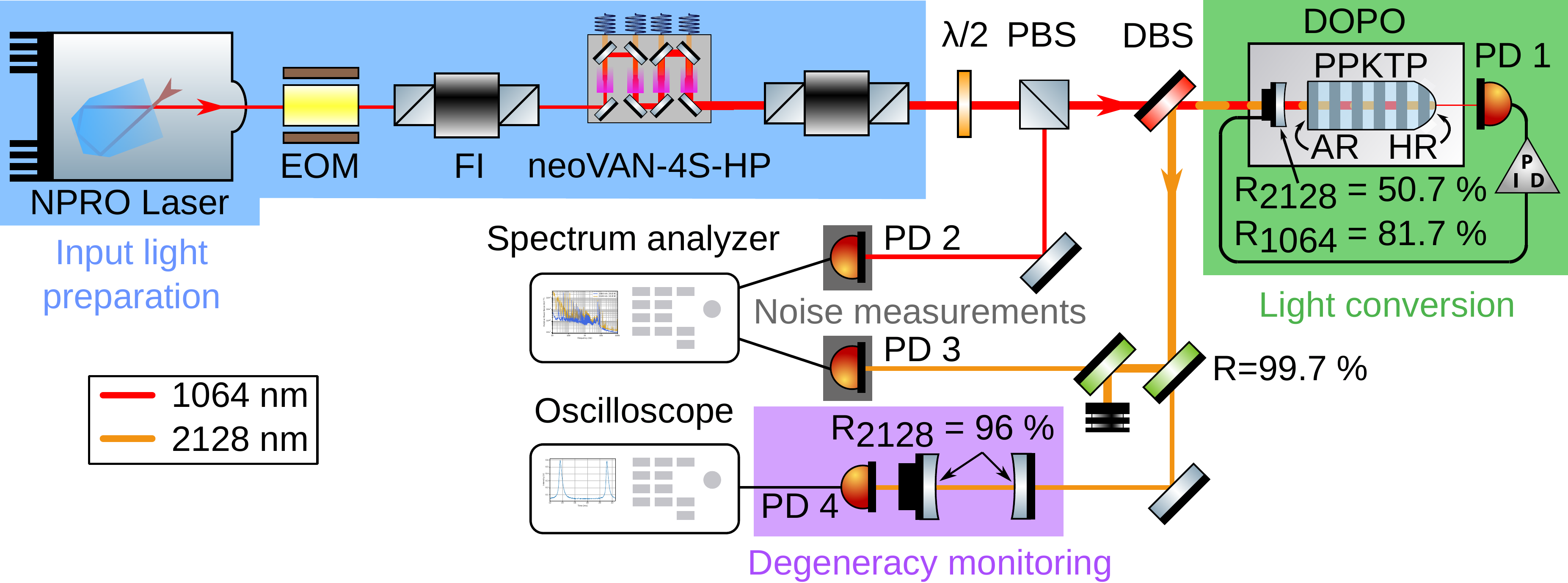}
	\caption{
		Schematics of the experiment.
		The NPRO laser and the neoVAN-4S-HP amplifier provided up to \SI{70}{W} output power at the wavelength \pump\,(coloured blue: Input light preparation) for the degenerate optical-parametric oscillator (DOPO) for wavelength doubling (coloured green: Light conversion).
		To ensure single frequency (degenerate) output, the converted light was monitored with a concentric resonator (coloured purple: Degeneracy monitoring).
		The relative power noise was detected with a photo diode and measured with a spectrum analyzer (coloured silver: Noise measurements).
		NPRO: non-planar ring oscillator laser; EOM: electro-optical modulator; FI: Faraday isolator; PBS: polarizing beam-splitter; DBS: dichroic beam-splitter; DOPO: degenerate optical parametric oscillator; PD: photo diode.
	}
	\label{fig:lasersystem}
\end{figure*}
\begin{figure}[t]
	\centering
	\includegraphics[width=\linewidth]{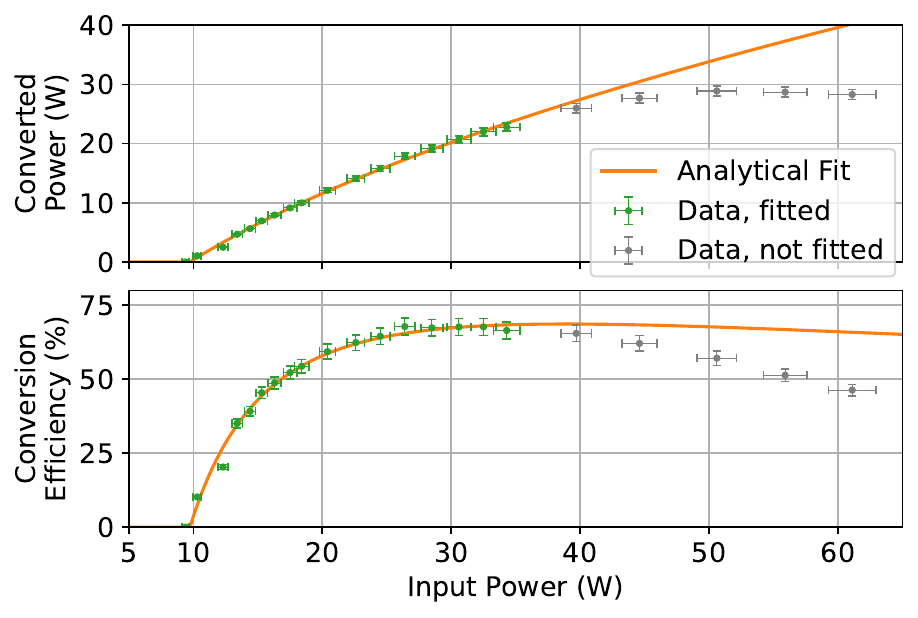}
	\caption{Power of the converted light at \convert{} (top) and external conversion efficiency (bottom) as a function of the input power at \pump.
		The indicated error bars correspond to the measurement accuracy of the thermal power meter, which was \SI{3}{\percent}. 
		The orange line shows a fit of the analytical formula for the converted power. 
		At the point of maximum conversion efficiency, $\SI{38.9 \pm 1.2}{\watt}$ of pump power was converted into $\SI{28.3 \pm 0.8}{\watt}$ of output power. 
		The maximum direct-measured converted light was $\SI{28.9\pm 0.9}{\watt}$.
		Above $\SI{34.3 \pm 1}{\watt}$ of input power, the error signal for the DOPO length control was distorted; therefore, those points were excluded from the fit (see text).}
	\label{fig:conversion}
\end{figure}
Here, we report the first realization of a stable, continuous-wave, high power laser system in the \SI{2}{\micro\metre} region, that meets GW observatory requirements. 
It uses high-efficiency degenerate optical parametric oscillation that is pumped by ultrastable light at \pump.
The latter is produced by a Nd:YAG nonplanar ring oscillator as the seed source, followed by a neoLASE amplifier providing up to \SI{70}{\watt}. 
A similar laser system is used in the currently operating Advanced LIGO detectors \cite{Bode2020}.
We directly observed $\SI{28.9\pm0.9}{\watt}$ of \convert{} light, a maximum external conversion efficiency of $\SI{72.8\pm 0.5}{\percent}$ and a relative power noise (around \SI{e-6}{\per\sqrt{\hertz}}), which was at Fourier frequencies above \SI{1}{\kilo\hertz} close to the \SI{1064}{\nano\metre} input light noise level.
While the motivation for this research is future gravitational-wave detectors, \SI{2}{\micro\metre} lasers are also employed in other applications, such as in the fields of material processing, free-space optical communication, gas sensing and as a potential laser sources for future atomic clocks \cite{JinXiaoxi2017, MINGAREEV20122095, Grady2010, HEMMING2014621, Robinson19, Robinson21}.

\section{Experimental Setup}\label{experimental-setup}
Our experimental setup (Fig.~\ref{fig:lasersystem}) was based on the first amplification stage of the laser system scheme as used in AdvancedLIGO for the fourth observation run O4 \cite{Bode2020}, followed by a nonlinear resonator that was optimised for wavelength-doubling via type 0 degenerate optical-parametric oscillation.
The seed laser was a \SI{2}{\watt} non-planar ring oscillator (NPRO) with a wavelength of \pump.
Its light was sent through an electro-optical modulator (EOM), which imprinted a phase modulation at \SI{28}{\mega\hertz} required for Pound-Drever-Hall locking of the nonlinear resonator.
A Faraday isolator (FI) protected the seed laser from back-reflections and back-scattering.
The light was then amplified by a neoLASE neoVAN-4S-HP laser amplifier up to power levels between 5 to 75 Watts \cite{Thies2019}.
A second Faraday isolator shielded the neoVAN amplifier from back-reflections.
A fraction of the high power beam was detected on a photodetector (PD 2) to perform  relative power noise (RPN) analysis, while the remaining light was used to pump the parametric process.
The degenerate optical parametric oscillator (DOPO) had a half-monolithic (hemilithic) design and was based on a periodically-poled potassium titanyl phosphate (PPKTP) crystal. 
A detailed description can be found in \cite{Darsow-Fromm2020, Darsow-Fromm2021}.
The highly-reflective coated, curved end face of the crystal formed one end of the resonator, while the other end of the crystal was anti-reflective coated for both wavelengths.
The resonator was completed by a separate coupling mirror with reflectivities of \SI{81.7}\% at \pump\ and \SI{50.7}\% at \convert{} (measured with a Agilent Cary 5000 spectrophotometer).
The parameters of the DOPO cavity are summarized in Table \ref{tab:cavity} for the pump and converted fields.
To stabilize the length of the DOPO on resonance, a modified Pound-Drever-Hall (PDH) control scheme was used in transmission of cavity together with a digital controller~\cite{darsow-frommNQontrolOpensourcePlatform2020} that fed back to the piezo-mounted coupling mirror (PD 1).
The parametric process converts the pump field into the signal and idler fields.
Here, we achieved degeneracy of these two fields by precisely controlling the temperature of two separate regions of the nonlinear crystal \cite{zielinskaFullyresonantTunableMonolithic2017, schonbeckaxelCompactSqueezedlightSource2018,Mehmet_2019} and monitored this state with a linear, concentric cavity (with mirror reflectivities of \SI{96}{\percent}; PD 4).
In our experiment long-term stability of the degenerate state could not be achieved; however, this is generally possible by temperature tuning, as shown in \cite{Darsow-Fromm2020}. 
The measurements presented here might include slightly non-degenerate operating points, which however had no discernible influence on the results.
A dichroic beam-splitter (DBS) separated the \pump\ and \convert\ light.
The RPN of the generated \convert\ output field was measured with an extended InGaAs photo diode (PD 3; Thorlabs FD05D) with custom transimpedance amplifier, and evaluated with a spectrum analyzer.

\section{Results}\label{results}
\begin{figure*}[t]
	\centering
	\includegraphics[width=0.9\linewidth]{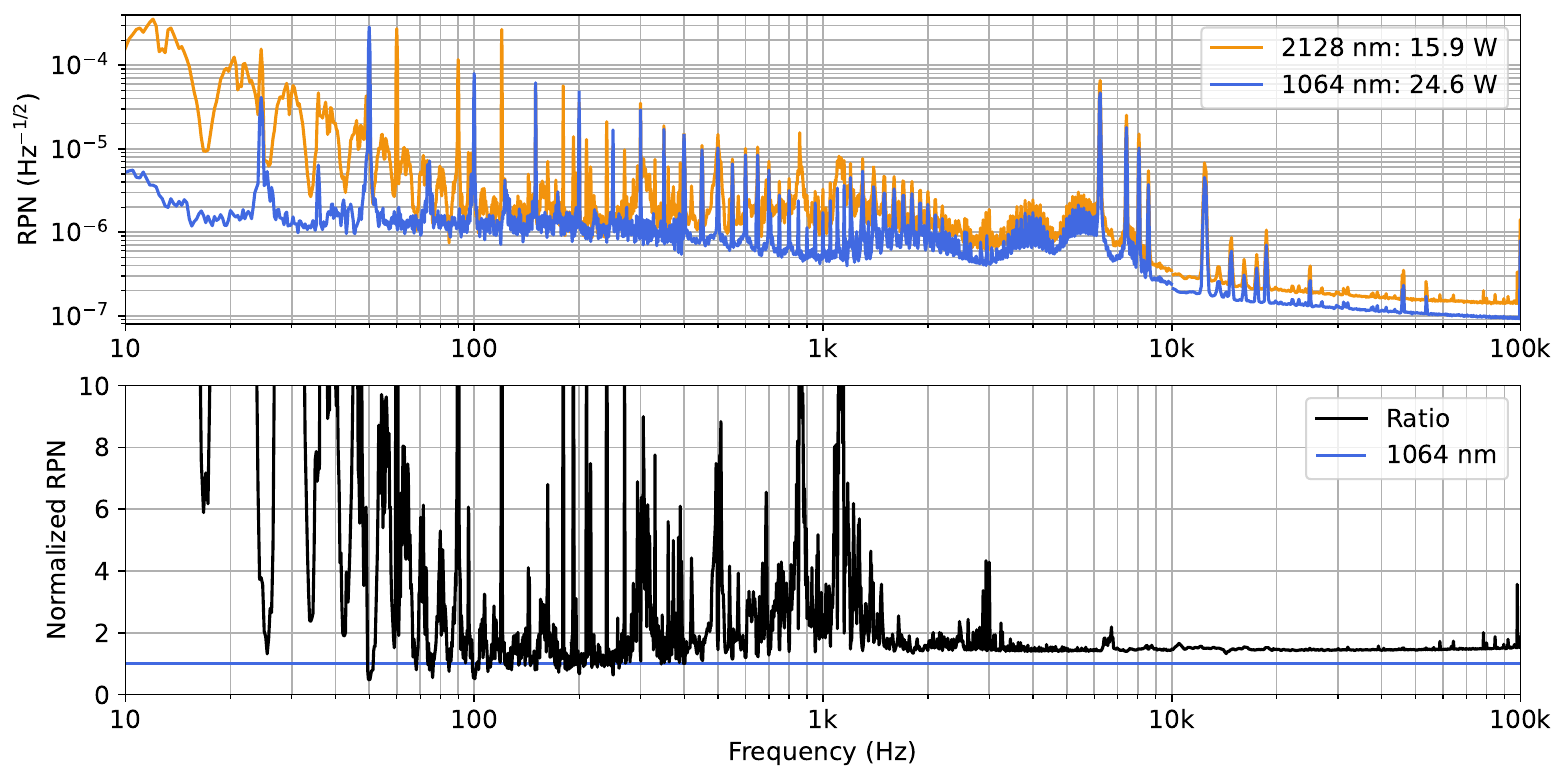}
	\caption{
		(Top) 
		Relative power noise (RPN) measurements of the \pump{} laser beam (blue) and the converted \convert{} laser beam (orange). 
		The relative power noise was somewhat higher than that of the input beam, however,	below \SI{100}{\hertz} completely dominated by detection noise from the extended InGaAs photo diode (Thorlabs FD05D). At higher frequencies, we measured only slightly increased relative noise. We suspect that this was caused by the positive slope of the conversion efficiency since the effect reduced when the conversion efficiency approached the theoretical, loss-limited
		maximum. (Bottom) RPN level of the converted beam (black) normalized to the RPN of the input (blue).}
	\label{fig:RPN}
\end{figure*}

\begin{table}
	\centering
	\caption{Overview of the degenerate optical parametric oscillator cavity parameters}
	\begin{tabular}{lrrl}
		\hline
		& 1064 nm         &  2128 nm &  \\
		\hline
		waist radius  & 31.5  &  44.9   & \textmu m \\
		finesse & 31.0   &  9.2   & \\
		free spectral range & 3.63  &  3.65  & GHz \\
		linewidth (FWHM) & 117  &  399   & MHz \\
		coupler reflectivity & 81.7 & 50.7 & \% \\
		power built-up & 19.2 & 5.9& \\
		\hline
	\end{tabular}
	\label{tab:cavity}
\end{table}
Fig. \ref{fig:conversion} shows the conversion efficiency $\eta$ as a function of the \pump{} input power $P_\text{in}$, as well as the measured \convert{} power $P_\text{out}$.
We did not correct $P_{out}$ or $\eta$ for power loss from imperfect mode matching, reflection loss of the crystal's AR coating, internal absorption, and residual transmission through the crystal's end surface coating.
An analytical fit has been made using the formula
\begin{align*}
P_\text{out} = 4\eta_\text{max}P_\text{th}\left(\sqrt{\frac{P_\text{in}}{P_\text{th}}}\right)\text{,}
\end{align*}
which is derived from \cite{breitenbach81ConversionEfficiency1995b, MMartinelli_2001}. The best fit was achieved for a threshold power $P_\text{th} = \SI{9.73\pm 0.12}{\watt}$, and the maximum efficiency $\eta_\text{max}= \SI{67.8 \pm 0.5}{\percent}$.
Corrected for the imperfect mode matching of $\SI{90.1\pm1.2}{\percent}$, the internal conversion efficiency was greater than \SI{75.3}{\percent}.
At the point of maximum efficiency, $\SI{38.9 \pm 1.2}{\watt}$ of pump power was converted into $\SI{28.3 \pm 0.8}{\watt}$ external light at \convert{}.
We obtained a maximum output power at \convert{} of $\SI{28.9 \pm 0.9}{\watt}$ at an input pump power of $\SI{50.6 \pm 1.5}{\watt}$, corresponding to an external conversion efficiency of still $\SI{57.1\pm 2.4}{\percent}$.
The uncertainty in these values are given by the \SI{3}{\percent} relative measurement error of our thermal power meter head, as specified by the manufacturer.
Fig. \ref{fig:RPN} shows the relative power noise in the input and output light beams, as simultaneously measured by two (extended) InGaAs photo diodes and a spectrum analyser. 
Electronic dark noise was negligible and therefore not subtracted.
From \SI{100}{\hertz} upwards the RPNs are similar. Any additional noise sources are probably electronic artefacts or mirror resonances coupling via pointing into RPN at the DOPO.
During the experiment, two effects were noted.
First, the DOPO PDH error signal became increasingly distorted with pump power, such that at a pump power of $\SI{30 \pm 0.9}{\watt}$ it was no longer possible to use this signal to control the resonator length electronically.
Instead, we opted to manually adjust the feedback voltage to hold the DOPO resonator on resonance.
Secondly, the conversion efficiency dropped quicker after the point of maximum conversion than expected from the theory.
We presume this was due to the imperfect length control of the resonator, combined with the increased thermal load on the resonator, which could have led to an onset of thermal lensing and a resulting decrease in mode matching.

\section{Conclusion}
Our experiment proves that a few tens of watts of ultra-stable light at \pump{} can be efficiently converted into light at \convert{} without massively increasing the relative power noise.
The required frequency degeneracy of the generated light by the optical parametric conversion process can be achieved by fine-tuning the temperature of the DOPO crystal.
Here, \SI{38.9\pm 1.2}{\watt} at \pump{} was converted to an external beam of \SI{28.9\pm0.9}{\watt} at \convert{}.
This is higher than the power level of \SI{3}{\watt} required for the low-frequency interferometer of the Einstein telescope \cite{ETSteeringCommitteeEditorialTeam2020}. 
The \SI{140}{\watt} needed for the future gravitational-wave observatory LIGO Voyager \cite{TheLIGOScientificCollaboration2022} seems to be reachable by using our shown procedure with further optimization of the DOPO parameters, as shown in \cite{Meier2010}, and potentially by using the technique of coherent beam combination \cite{Wellmann2021} from additional sources.

\begin{backmatter}
\bmsection{Funding} This research has been funded by the Germany Federal Ministry of Education and Research, grant no. 05A20GU5 and Deutsche Forschungsgemeinschaft (DFG, German Research Foundation) under Germany’s Excellence Strategy – EXC-2123 QuantumFrontiers – 390837967.

\bmsection{Disclosures} The authors declare no conflicts of interest.

\bigskip
\noindent

\end{backmatter}

\bibliography{references}

\end{document}